\documentclass[11pt]{article}

\usepackage[margin=.75in]{geometry}
\usepackage{multirow}
\usepackage{color}
\usepackage{amsmath, amsthm, amssymb,fancyhdr,mathrsfs,  enumerate}
\usepackage{graphicx}
\usepackage{latexsym}
\usepackage[round, authoryear]{natbib}

\newcommand{\mbold}[1]{\mbox{\boldmath $#1$}}
\newcommand{\lo}{{\bf\bar\omega}}
\newcommand{\cc}{\textcolor{black}}
\newcommand{\rr}{\textcolor{black}}
\newcommand{\jj}{\textcolor{black}}
\newcommand{\ww}{\textcolor{black}}
\newcommand{\BL}{\textcolor{black}}

\begin{document}

\begin{center}
      \Large{\bf {Proportion estimation based on a   partially rank ordered set sample   with   multiple concomitants in a  breast cancer study}}
\end{center}
\begin{center}
 {\sc Armin Hatefi} \footnote{  { Corresponding author}\\
              Email: Hatefi.ar@gmail.com \& armin.hatefi@umanitoba.ca}
 and {\sc Mohammad Jafari Jozani} 

 Department of Statistics,  University of Manitoba, Winnipeg, MB, Canada, R3T 2N2. 
             
\end{center}
\begin{center} {\small \bf Abstract}: 
\end{center}
In this paper, we  use   partially rank-ordered set (PROS) sampling design  with multiple concomitants  in a  breast cancer study  and propose a method  to   estimate the  proportion of patients with  malignant (cancerous) breast   tumours in a given population.
  Through extensive numerical studies, the performance of the estimator is evaluated under  various concomitants with different ranking potentials (i.e., good, intermediate and bad) and  tie-structures. We  show  that the  PROS  estimator with multiple concomitants   based on the  ranking information provided through  some easy to obtain cytological characteristics  that are associated with the malignancy of breast tumours performs  better than its  counterparts under simple random sampling (SRS) and ranked set sampling (RSS) designs with logistic regression models.  As opposed to available RSS based methods in the literature, our proposed methodology   allows  to declare ties among the ranks and does not rely on the existance of  any specific regression model assumptions.



\section{ Introduction } \label{sec:intro}
In many medical studies, measuring the variable of interest (e.g., disease status) is   difficult and   involves  complicated procedures that are usually   time consuming   and/or  expensive.  However, one  may have access to several  concomitant variables (e.g., laboratory and demographic characteristics) from the sampling units  that  can  be quantified easily   at little cost.  In most applications, 
 the auxiliary information is often  used  in the estimation   process  to make better inference about  the parameter  of interest. 
For example, in a diabetes study,  \ww{ \cite{chen2005ranked} used the  body mass index, weight and 
 buttocks circumference to provide efficient estimate of  the diabetes status of  patients.} 
Or, in a cardiovascular disease study,  \ww{ due to  the association between the smoking status,  the body mass index, the dietary beta carotene  intake and beta plasma concentration in blood (as the quantitative trait of interest),  \cite{schlattmann2009medical} used  these concomitants  to better predict   the incidents of  cardiovascular diseases.} 
   In   such examples,  
   the sampling units can  be ranked easily    (at little cost)  using the available concomitants  prior to taking the final measurements on the variable of interest. 
   In these settings one can use  the  information of these concomitants  in the    data  collection process of the study to obtain more representative samples from the underlying population.  Rank-based sampling designs such as ranked set sampling (RSS) and partially rank ordered set (PROS) sampling provide a collection of  techniques to obtain and analyze these expensive measurements with the help of inexpensive information. 
 
 In this paper, we  use the PROS sampling design with multiple concomitants in a breast cancer study and  propose a methodology that can be used to better estimate the proportion of patients with  cancerous tumours compared with the commonly used  methodologies  in the literature based on simple random sampling (SRS) and RSS designs.  In  breast cancer studies, the discrimination between malignant (cancerous)  and benign (not  cancerous)  tumours is very important and  requires a comprehensive biopsy procedure. Benign tumours are those  that  cannot spread to other parts of body   and only  grow locally.  In contrast, malignant tumours are those    that  invade and destroy nearby tissues and spread to other parts of the body.  To determine whether tumours are malignant or benign, samples of suspected masses  are either  surgically biopsied or  verified by clinical re-examinations after  3 to 12 months following  the aspiration of breast clumps.   This  is an invasive procedure that involves the physical extraction of tissue. The test is expensive, and the results tend to take some times  to process.  
  To accurately diagnose  breast cancer  based  purely on a  Fine Needle Aspiration (FNA),
   \ww{\cite{wolberg1990multisurface} identified  nine visually  assessed cytological characteristics of an FNA sample in the  Wisconsin Breast Cancer Data (WBCD) that are considered to be relevant to breast cancer determination and its diagnosis.} 
 As we explain in Section 2, these cytological characteristics  (concomitants)  are easy to obtain and can play an important role in  the early  discrimination between malignant and benign breast masses.  
More references pertaining to  WBCD \cc{can be} found in \ww{ \cite{wolberg1990multisurface}} and  \ww{\cite{terpstra2004concomitant}} 
as well as  the website of the  data set \ww{(\citealp{Bache+Lichman:2013})}
and references therein.

%

Several studies have been done on the WBCD to show  the benefits of using  rank-based sampling techniques for  breast cancer research.  For example, 
 \ww{\cite{terpstra2004concomitant}} proposed an RSS-based methodology to estimate  the probability of having   malignant breast  tumours  and the proportion of patients with breast cancer  in  the WBCD   when only  a single concomitant is used in the  data collection  process. 
Although they proposed a more efficient estimator based on RSS data for the  population proportion, their statistical methodology   is only applicable in bivariate settings. In other words,   one can take advantage of one and only one concomitant in the estimation of the population proportion.  
In  the  WBCD,  however,   there are  multiple  concomitants  that are highly correlated with  the malignancy of breast tumours and one  might want  to incorporate them  simultaneously in to the estimation as well as the data collection  process  to improve the inference about  the  proportion of patients with  malignant  breast tumours. 
To this end, \ww{\cite{chen2005ranked}} proposed an RSS-based  methodology using multiple concomitants to estimate the population proportion.  However, their method is based on  the  logistic regression modeling assumption and, as we show in Section \ref{sec:4},   even  by  using  concomitants   that are highly correlated with the response variable,  the logistic regression model might not be   statistically significant. This could  be because in these applications the sample sizes are not necessary high. In the absence of such strong modeling assumption, it is not clear how the extra information should be incorporated in the  estimation process  in a more efficient way.
  In addition, \ww{the regression-based methodology  requires  training samples  with measurements on both  concomitants and  the response variable  to estimate  the  parameters of the  model. In many applications, however, training samples  are not available.}
Morover, none of the  \ww{one-concomitant RSS (\citealp{terpstra2004concomitant}) and  RSS-based logistic regression  methods (\citealp{chen2005ranked})  allows  to declare  ties among the  ranks in the ranking process of the  RSS  scheme.} This  is not   realistic,    especially for  the WBCD   where all the cytological concomitants are \BL{(categorical) ordinal variables} taking on values between 1 to 10.  
  In such a case,   forcing rankers to assign unique ranks to  the sampling units results in  random assignment of the ranks and can  lead to substantial amount of  ranking error and consequently to poor/invalid statistical inference. 
  
  To overcome these problems,  in this paper, \ww{we propose to use  the partially rank ordered set (PROS) sampling technique (\citealp{ozturk2011sampling}) with  
   multiple concomitants and construct   more  efficient estimate of the population    proportion and apply it to the  WBCD.} 
\BL{In the PROS sampling technique, one  can assign sampling  units into partially rank-ordered subsets, instead of precisely ranking all sampling units in a set.
Let $G=\{g_1,\ldots,g_l\}$ denote a partition of the integers $\{1,\ldots,H\}$ into $l$ mutually exclusive subsets $g_r=\{(r-1)m+1,\ldots,rm\}$,   each of size $m$, where  $g_r=\{(r-1)m+1,\ldots,rm\}$ and $m=H/l$.
To construct a PROS sample from the population of interest, a set of $H$ units are first selected.  Instead of ranking all units in the set,  a ranker   is asked to assign the sampling units into   subsets  that are partially rank-ordered so that units in subset $g_r$ have smaller ranks than units in subset $g_s$, when $s > r$ 
; $r,s=1,\ldots,l$. This can be done using   concomitant variables or any other means that does not require measuring  the variable of interest. 
 From the subset $g_1$, a unit is selected at random for full measurement, say  $Y_{[g_1]1}$. Selecting another $H$ units and again assigning them into subsets, we select  a unit at random from the subset $g_2$ for full measurement, namely $Y_{[g_2]1}$. The process is continued until we obtain $Y_{[g_l]1}$ as the final measurement from the subset $g_l$. This results in a PROS sample of size $l$  as $Y_{[g_1]1},\ldots,Y_{[g_l]1}$ and the whole process constitutes  one cycle of the sampling procedure.  The cycle can be repeated $n$ times to generate a PROS sample of size $nl$, denoted by $\{Y_{[g_r]i};r=1,\dots,l;i=1,\ldots,n\}$.  In the above description, the number of observations in each  subset is assumed to be the same.  However, one can easily relax this assumption with slight modifications in the notations.\  In this paper, as we  illustrate in Subsection \ref{sub:mcp}, we consider a general setting where we do not assume that  the number of subsets and subset sizes  are fixed through the sampling procedure.\ This is because, in practice,  the number of subsets are  actually determined by  concomitants  or the ranking ability of rankers.  }

\begin{table}
\begin{center}
\BL{\caption{An example of PROS sample}
\begin{tabular}{cccc} \hline\hline
Cycle & Set &  Subsets & Observation \\ \hline
 1    & ${\bf U}_1$ & $G_1=\{\mbold{g_1},g_2,g_3\}=\{ \mbold{\{1,2\}},\{3,4\},\{5,6\} \}$ & $Y_{[g_1]1}$  \\ 
      & ${\bf U}_2$ & $G_2=\{g_1,\mbold{g_2},g_3\}=\{\{1,2\} ,\mbold{\{3,4\}},\{5,6\} \}$ & $Y_{[g_2]1}$  \\
      & ${\bf U}_3$ & $G_3=\{g_1,g_2,\mbold{g_3}\}=\{ \{1,2\},\{3,4\},\mbold{\{5,6\}} \}$ & $Y_{[g_3]1}$  \\ \hline
 2    & ${\bf U}_1$ & $G_1=\{\mbold{g_1},g_2,g_3\}=\{ \mbold{\{1,2\}},\{3,4\},\{5,6\} \}$ & $Y_{[g_1]2}$  \\
      & ${\bf U}_2$ & $G_2=\{g_1,\mbold{g_2},g_3\}=\{ \{1,2\}, \mbold{\{3,4\}},\{5,6\} \}$ & $Y_{[g_2]2}$  \\
     & ${\bf U}_3$ & $G_2=\{g_1,g_2,\mbold{g_3}\}=\{ \{1,2\},\{3,4\},\mbold{\{5,6\}} \}$ & $Y_{[g_3]2}$ \\ \hline
 \end{tabular}}
  \label{ta:pros}
 \end{center}  
 \end{table}

\BL{Table \ref{ta:pros} illustrates  a simple example of the construction of the  PROS sampling design when $H=6$, $l=3$, $m=2$, the  cycle size is $n=2$ and $G=\{g_1,g_2,g_3\}=\{{\{1,2\}},\{3,4\},\{5,6\}\}$. Each set contains six units that  are assigned to three partially rank-ordered  subsets using  a ranking mechanism.  The partial ranking indicates that ties have been declared among the ranks for the units within subsets; however, the subsets themselves are partially ordered.  
More specifically,  subsets $g_1$ include the units with the two smallest judgment ranks among the six units. Units in subsets $g_2$ have judgment ranks greater than units in subsets $g_1$ and smaller judgment ranks than units in subset $g_3$. Units in subsets $g_3$ have received  the two highest judgment ranks among the units in each set. In each set, one of the units is finally selected for full measurement from the bold faced subsets in Table \ref{ta:pros}. The resulted PROS sample of size $nl=6$ is then  denoted  $\{Y_{[g_r]i},r=1,2,3;i=1,2\}$.
 } 
   
Recently, the PROS sampling design has received considerable attention in the literature.  For example, 
\ww{\cite{hatefi2014information}} studied the information and uncertainty structures of PROS data. 
\ww{\cite{ozturk2014inclusion}} used  PROS samples for estimation problems in finite population
settings. \ww{\cite{nazari2014}} developed nonparametric kernel density estimators
using PROS data. \ww{\cite{hatefi2013mixture}} applied PROS sampling in mixture modeling to estimate the age structures of short-lived fish species. \ww{\cite{ozturk2013combining}} and \ww{\cite{frey2012nonparametric}} relaxed the assumption concerning the pre-specification of the number of subsets in each set. Due to different ranking potentials of the concomitants, the results of \ww{\cite{ozturk2013combining}} and \ww{\cite{frey2012nonparametric}}  allow to  declare as many subsets as desired for the accommodation of  tied ranks among the units in the sets. 
\ww{\cite{ozturk2013combining}} also studied the statistical inference based on  multi-observer RSS  in the estimation of the  population mean. Recently, \ww{\cite{ozturk2013estimation}} used the properties of PROS samples under multiple auxiliary information in the estimation of the population mean and total in finite population settings.

 In this study, using multiple concomitants, we  first explore the benefits of  employing   PROS sampling technique   for estimating  the population proportion. Then,  the  estimation procedure  will be  applied  to the  WBCD to provide  more accurate  estimate  of the proportion of patients with  malignant breast  tumours. 
To this end, Section \ref{sec:2} describes PROS sampling design using multiple concomitants for analysis of the WBCD. We propose an estimation procedure using multi-concomitant PROS samples for  the population proportion in Section \ref{sec:3}.  In Section \ref{sec:4}, through two numerical analyses, we study the performance of the multi-concomitant PROS procedure in the estimation of the proportion of patients with  malignant breast   tumours   compared with those based on one-concomitant RSS and multi-concomitant RSS-based logistic regression methods.  Summary and concluding remarks are finally presented in Section \ref{sec:5}. 

\section{Multi-concomitant based sampling from the  WBCD} \label{sec:2}
 Breast cancer accounts for one of the most important types of cancers in women and  causes a significant rate of death worldwide. 
 Most of  breast cancer cases are discovered when a noticeable lump feels differently from the rest of breast tissues. The  early detection  of cancerous tumours  is  very important  in the treatment of breast cancer. \ww{The earlier breast cancer is detected, the better it may be for the patientÕs long-term health. Many breast cancer organizations, such as the American Breast Canter Foundation, the Cancer Research UK and the Canadian Breast Cancer Foundation, are concerned about the incidence rate and prevalence of breast cancers in target populations at given times.}  
To this end, regular studies are conducted to estimate the prevalence of malignant breast cancer tumours among the patients in the target population. This is usually done by  estimating the proportion of patients with a  new or previous  malignant breast tumours. 
In this paper,  we  consider  the WBCD  data set as our population   and show that  PROS sampling design using multiple concomitants  can be used as an efficient tool for estimating  the proportion of \cc{patients} in the population with malignant breast  tumours.   This data set was collected by Dr.\ William H.\ Wolberg (Department of Surgery, University of Wisconsin, Madison) and is available   \ww{online at the UCI  machine learning repository (\citealp{Bache+Lichman:2013})}.

Suppose  the dichotomous  variable $Y$ 
(hereafter called  the \BL{Malignant Tumours}) 
denotes \BL{the status of breast masses} as malignant (success) or benign (failure) tumour. 
The malignancy of the breast tumours is determined through a comprehensive biopsy procedure. To accurately diagnose the breast tumour samples based purely on FNA, Dr.\ Wolberg identified nine visually assessed characteristics of an FNA sample and exploited them to determine  the status of the tumour samples and start diagnosing  them with proper procedures. To be more specific, assessing the epithelial cell clumps obtained through an standard method of breast FNAs,
\ww{\cite{wolberg1990multisurface}  first  identified eleven cytological characteristics of FNAs to distinguish between benign and malignant tumours.} 
These eleven cytological characteristics \BL{of breast FNAs were valued} on a scale of 1 (normal) to 10 (most abnormal) \BL{ by a doctor 
assessing the tissue cells through a microscope,}
such that 1 indicates the closest status to the benign while 10 represents the most anaplastic.$\BL{^{\footnotesize{3, 4}}}$ Statistical analysis found that nine of these  cytological characteristics  play significant roles in the discrimination between malignant and benign breast masses. These nine cytological characteristics are as follows: the amount of thickness (Clump Thickness),  the  surrounding cells cohesion (Marginal Adhesion) of the epithelial cell aggregates, the size of an epithelial cell aggregate (Single Epithelial Cell Size) calculated as the diameter of the population of the largest epithelial cells relative to erythrocytes, the proportion of  a single epithelial nuclei being bare of the peripheral cytoplasm (Bare Nuclei), the blandness of the nuclear chromatin (Bland Chromatin), the normality of nucleolus (Normal Nucleoli), the unusual mitoses (Mitoses), the uniformity in  size (Uniformity of Cell Size) and  the shape (Uniformity of Cell Shape) of the epithelial cell.


%
{\small
\begin{table}[h!]
\caption{\small{Cytological concomitants and their correlations with \BL{Malignant Tumours} in the WBCD.}}
\vspace{0.3cm} 
\centering 
\small{\begin{tabular}{lc|lc}\hline\hline
Concomitants             & $\rho$  &  Concomitants               & $\rho$   \\\hline
Bare Nuclei              & 0.8226  & Marginal Adhesion           & 0.7062    \\
Uniformity of Cell Shape & 0.8218  & Single Epithelial Cell Size & 0.6909   \\
Uniformity of Cell Size  & 0.8208  & Mitoses                     & 0.4234   \\
Bland Chromatin          & 0.7582  & \BL{Subject ID}             & \BL{-0.0847}  \\
Normal Nucleoli          & 0.7186  & \BL{Independent Covariate}  & \BL{-0.0317}  \\
Clump Thickness          & 0.7147  &                             &          \\\hline
\end{tabular}
\label{tab:cor}}
\end{table}
}


We use  these nine easy to obtain  cytological characteristics as concomitants associated with  
\BL{malignancy of the corresponding FNA sample to conduct the  PROS sampling designs with multiple concomitants from the WBCD}. Table \ref{tab:cor} shows these concomitants and their correlations with the  \BL{Malignant Tumours}. 
In addition to these nine cytological concomitants, to explore the effect of  unreliable concomitants (i.e., those with very low  correlations with the \BL{Malignant Tumours}), we treat the \BL{Subject ID} (the unique code for each subject) as another concomitant  in our numerical studies. Due to the nature of this concomitant, it is seen from Table \ref{tab:cor} that its correlation with the \BL{Malignant Tumours} is $-0.0847$.
\BL{Also, to investigate  the effect of an independent concomitant on the performance of our proposed methodology, we generated an independent ordinal  variable, namely ``Independent Covariate'', taking values on 1 to 10. This guarantees almost zero correlation between the Independent Covariate and the Malignant Tumours. }
\subsection{Multi-concomitant PROS sampling}\label{sub:mcp}

Let $({\bf X},Y)^{\top}$ denote a multivariate random variable when  $Y$ represents  the  \BL{response} variable following a Bernoulli distribution with parameter $p$ (i.e., probability of success) and  ${\bf X}=(X_1,X_2,\ldots,X_l);\, l \ge 1$  \cc{denotes} the  $l$-variate   concomitants.
In order to construct a  PROS  sample with  multiple concomitants (hereafter called multi-concomitant PROS),
we first  specify two positive integers $H$ as the  set size and $\cc{n}$ as the  cycle size. 
Throughout the paper, we 
 assume that  $N=nH$, where $N$ represents the total sample size and  the number of observations from each   judgment order class is $n$.    We  now describe how to obtain a multi-concomitant PROS sample of size $N$, while a numerical illustration  is provided in   Section 2.2.   To measure the $r$-th judgment order statistic, say $Y_{[r]i}~ (r=1,\ldots,H; i=1,\ldots,n)$ through this design, we randomly choose a set of $H$ experimental units,  ${\bf U}^{[r]}_{i}=\{u_{1i},\ldots,u_{Hi}\}$  from the population.
 Let $K$ concomitants be available for  the ranking purpose and suppose   ${\bf X}_{k,i}^{[r]}=(X_{1,k,i},\ldots,X_{H,k,i})$ represents the values of the $k$-th concomitant, say $X_k$, $ k=1,\ldots,K$,  for the sampling  units in  ${\bf U}^{[r]}_{i}$. After ranking the sampling units using  ${\bf X}_{k,i}^{[r]}$, we construct the rank vectors as follows
 \[
 {\bf O}_{k,i}^{[r]}=O(X_{1,k,i},\ldots,X_{H,k,i})=\{O_{1,k,i},\ldots,O_{H,k,i}\}, ~~~ k=1,\ldots,K,
 \] 
 where $O_{h,k,i}$ is the rank assigned to the unit $u_{hi}\in {\bf U}^{[r]}_{i}$.
  If ranking is done based on   \BL{an ordinal} variable (e.g., the cytological characteristics), the tied ranks 
 may be produced. In this situation, all tied units in the set receive the same rank. If there is a negative correlation between the concomitant variable and the  \BL{response} variable $Y$, the ranking operator is selected as $O_{r,k,i}=H+1-O_{H+1-r,k,i}, \, r=1,\ldots,H,$ to produce the necessary judgment order statistics. Note that ${\bf O}_{k,i}^{[r]}$ is the rank vector associated with units in the set ${\bf U}^{[r]}_{i}$ from which we derive the $r$-th judgement order statistic, $Y_{[r]i}$.  
For each ${\bf O}_{k,i}^{[r]}, k=1,\ldots,K;i=1,\ldots,n$, we build  an  $H \times H$ weight matrix, \cc{ ${\bf D}^{[r]}_{k,i}$}, whose  rows and columns stand for the units of the set ${\bf U}^{[r]}_{i}$ and assigned judgment ranks, respectively. The entities of the \cc{ ${\bf D}^{[r]}_{k,i}$} stand for the strength-of-weights of the ranking procedure. If there is no tie in the ranking vector ${\bf O}_{k,i}^{[r]}$, the $h$-th $(h=1,\ldots,H)$ row and the $O_{r,k,i}$ th column of the matrix \cc{ ${\bf D}^{[r]}_{k,i}$} is one, and other entries of the $h$-th row is zero. If there are $m$ tied ranks for the $h$-th unit ($u_{hi}$), then all the entries corresponding to the tied ranks in the $h$-th row will be $1/m$ and other entries of the row will be zero. In a similar fashion, we build   \cc{ ${\bf D}^{[r]}_{k,i}$} for all $k=1,\ldots,K$. To incorporate the ranking information obtained from all  available concomitants in the selection of  $Y_{[r]i}$, we focus on the weighted mean of the strength-of-weight matrices
\[
{\bf\bar D}^{[r]}_i= \sum_{k=1}^{K} \alpha_{k} \, \cc{ {\bf D}^{[r]}_{k,i}},
\]
where $\sum_{k=1}^{K} \alpha_k=1$. The coefficients $\alpha_k$ is chosen to reflect the importance of the concomitant variable $X_k$ in the ranking process. \ww{Due to \cite{lynne1977ranked}, it is known that  larger values for  the  correlation coefficient between a concomitant and \BL{the response} variable results in a better precision for the corresponding  concomitant-based RSS estimator.} As proposed by \ww{\cite{ozturk2013estimation}}, we calculate 
\[
\alpha_k=\frac{|\rho_k|}{\sum_{k=1}^{K} |\rho_k|},
\]
where $\rho_k$ denotes the correlation coefficient  between $Y$  and $X_k,k=1,\ldots,K$. 
Finally $Y_{[r]i}$  will be obtained from the  unit having the maximum entry in the $r$-th column of ${ \bf\bar D}^{[r]}_{i}$. If the maximum is unique, say $s$, we observe  $(Y_{[r]i},\lo_i^{[r]})$ as the multi-concomitant PROS sample, where $Y_{[r]i}$ is the \BL{response} variable of the unit $u_{si}$ in the set  ${\bf U}_i^{[r]}$ and $\lo_i^{[r]}$ will be the $s$-th row of the ${\bf\bar D}^{[r]}_{i}$.
However, if the maximum entry in the $r$-th column of ${\bf\bar D}^{[r]}_{i}$ is not unique, we consider all the rows with the  maximum weight. For all these rows, we calculate the concentration around $r$-th judgment order statistic. Then the $Y$-measurement of the unit with the highest concentration is identified as $Y_{[r]i}$. If the highest concentration is still not unique, one of the units is randomly selected for full measurement and we calculate its corresponding weight vector $\lo_i^{[r]}$. For a given weight vector $\lo_i^{[r]}=({\bar\omega}_i^{[r,1]},\ldots,{\bar\omega}_i^{[r,H]})\in {\bf\bar D}^{[r]}_{i}$, we calculate the concentration around the $r$-th judgment order statistic as 
\begin{eqnarray*}
\gamma_{r,i}=\sum_{t=1}^{H} (t-r)^2 {\bar\omega}_i^{[r,t]}. 
\end{eqnarray*}
Small values of $\gamma_{r,i}$ imply a  high concentration around  the $r$-judgement rank. 
Finally the observed multi-concomitant PROS data are given by 
\[
\left\{ (Y_{[r]i},{\lo}^{[r]}_i), ~ r=1,\ldots,H;i=1,\ldots,\cc{n}\right\}.
\]
Note that through multi-concomitant PROS sampling, we not only  measure $Y_{[r]i}$ but also we  calculate  the weight vectors  ${\lo}^{[r]}_{i}$ related to the judgment ranks using combined ranking information from all the available concomitants. Moreover, $Y_{[r]i}$ and ${\lo}^{[r]}_i$ are dependent random variables for fixed $i$ and $r$.
In a similar vein, the PROS sampling based on $K$ concomitants can be extended  to the PROS sampling under $K$ different rankers for only one concomitant.
See \ww{ \cite{ozturk2013combining} for more details}.

\subsection{An illustrative example}

To illustrate the construction of  multi-concomitants PROS samples,  we present  a simple example based on the  WBCD using $K=4$ cytological characteristics as concomitants  including the \BL{Subject ID} $(k=1)$, Uniformity of Cell Size $(k=2)$, Uniformity of Cell Shape $(k=3)$ and Bare Nuclei $(k=4)$. Suppose we are interested in measuring the second judgment order statistic $Y_{[2]1}$ when the  set size $H=5$ and the cycle size $n=1$. 
{\small
\begin{table}[h]
\caption{\small{Units of the set ${\bf U}^{[2]}_1$ and their concomitants in the illustrative  example.}}
\vspace{0.3cm} 
\centering 
\small{\begin{tabular}{ccccc}\hline\hline
   & \multicolumn{4}{c}{Concomitants} \\ 
                  \cline{2-5} 
Units      &  \BL{Subject ID}    & Uniformity of Cell Size  & Uniformity of Cell Shape &  Bare Nuclei  \\ 
                &    $k=1$                     &    $k=2$                           &   $k=3$                                & $k=4$             \\ \hline
$u_{11}$   &  1033078        & 1          &1           &1   \\
$u_{21}$   &  1035283        & 1          &1           &1    \\
$u_{31}$   &  1016277        & 8          &8           &4       \\
$u_{41}$   &  1017122        & 10         &10          &10       \\
$u_{51}$   &  1044572        & 7          & 5          & 9 \\\hline
\end{tabular}
\label{tab:ex1}}
\end{table}
}

Table \ref{tab:ex1} shows a set of five units ${\bf U}^{[2]}_1=\{u_{11},u_{21},u_{31},u_{41},u_{51}\}$ and their concomitants selected for ranking process in this example.
To illustrate the effect of tie-structure in the example, we assume that  tied-ranks may be declared in the ranking process  using  three cytological characteristics (i.e., $k\in\{2, 3 ,4\}$).  Unique ranks are  also assigned to the units using the   \BL{Subject ID}.
 The tie-structures in this example are constructed as follow.  The values of concomitants  are assigned to the units in subsets  $s_1=\{1,2\},s_2=\{3,4\},s_3=\{5,6\},s_4=\{7,8\}$ and $s_5=\{9,10\}$ and these subsets \cc{are} partially ranked so that units  within each group receive the same rank and units  in $s_h$ receive lower ranks than units  in $s_{h^{'}}$ when $h<h^{'}$. 
For example, using the Bare Nuclei values reported in Table \ref{tab:ex1} for the judgment ranking process, 
units $u_{11},u_{21}$ are declared tied at ranks $\{1,2\}$, unit $u_{31}$ is uniquely ranked $3$ and units $u_{41},u_{51}$ receive tied ranks at $\{4,5\}$.
{\small
\begin{table}[h]
\caption{\small{The tie-structures of the set ${\bf U}^{[2]}_1$ and the ranks declared in the curly brackets.}}
\vspace{0.3cm} 
\centering 
\small{\begin{tabular}{cccccc}\hline\hline
Concomitants   & \multicolumn{5}{c}{Units} \\ 
                  \cline{2-6} 
$k$   &    $u_{11}$     & $u_{21}$      & $u_{31}$     &  $u_{41}$   & $u_{51}$   \\ \hline
1     &    1,~\{3\}     & 1,~\{4\}      & 1,~\{1\}     & 1,~\{2\}    & 1,~\{5\}   \\
2     &    1/2,~\{1,2\} & 1/2,~\{1,2\}  & 1/2,~\{3,4\} & 1,~\{5\}    & 1/2,~\{3,4\}     \\
3     &    1/2,~\{1,2\} & 1/2,~\{1,2\}  & 1,~\{4\}     & 1,~\{5\}    & 1,~\{3\}     \\
4     &    1/2,~\{1,2\} & 1/2,~\{1,2\}  & 1,~\{3\}     & 1/2,~\{4,5\}& 1/2,~\{4,5\}     \\\hline
\end{tabular}
\label{tab:ex2}}
\end{table}
}

Table \ref{tab:ex2} provides the tie-structures and ranks declared for the units of the set  ${\bf U}^{[2]}_1$. Under the Bare Nuclei, for instance,
 the weights  are $1/2$ for the $u_{11},u_{21}$ tie-ranked at $\{1,2\}$, $1$ for $u_{31}$ ranked $3$ and $1/2$ for units $u_{41},u_{51}$ tie-ranked at $\{4,5\}$. Using the  \BL{Subject ID}, as noted earlier, we observe  unique ranks (i.e., no tied-rank) for all the units in the first row of Table \ref{tab:ex2}.
From Table \ref{tab:ex2}, we construct the weight matrix \cc{${\bf D}^{[2]}_{k,1}$}, $k=1, \ldots, 4$,  based on these  concomitants  as follow 
\begin{align*}
\cc{{\bf D}^{[2]}_{1, 1}}&=\left[\begin{array}{ccccc} 
                0 & 0 & 1 & 0 & 0\\
                0 & 0 & 0 & 1 & 0\\
                1 & 0 & 0 & 0 & 0\\
                0 & 1 & 0 & 0 & 0\\
                0 & 0 & 0 & 0 & 1\\
\end{array} \right], \quad 
~\cc{{\bf D}^{[2]}_{2, 1}}=\left[\begin{array}{ccccc} 
                \frac{1}{2} & \frac{1}{2} & 0 & 0 & 0\\
                \frac{1}{2} & \frac{1}{2} & 0 & 0 & 0\\
                0           & 0           &  \frac{1}{2} & \frac{1}{2} & 0\\
                0           & 0           & 0 & 0 & 1\\
                0           & 0           &  \frac{1}{2} & \frac{1}{2} & 0\\
\end{array} \right],\\
\cc{{\bf D}^{[2]}_{3, 1}}&=\left[\begin{array}{ccccc} 
                \frac{1}{2} & \frac{1}{2} & 0 & 0 & 0\\
                \frac{1}{2} & \frac{1}{2} & 0 & 0 & 0\\
                0           & 0           & 0 & 1 & 0\\
                0           & 0           & 0 & 0 & 1\\
                0           & 0 & 1 & 0   & 0\\
\end{array} \right],\quad
\cc{{\bf D}^{[2]}_{4, 1}}=\left[\begin{array}{ccccc} 
                \frac{1}{2} & \frac{1}{2} & 0 & 0 & 0\\
                \frac{1}{2} & \frac{1}{2} & 0 & 0 & 0\\
                0           & 0 & 1 & 0   & 0        \\
                0           & 0           & 0 & \frac{1}{2} & \frac{1}{2}\\
                0           & 0           & 0 & \frac{1}{2} & \frac{1}{2}\\
\end{array} \right].
\end{align*}
Using ${\bf \alpha}=(0.0468, 0.0453, 0.4537, 0.4542)$ \cc{obtained based on the} correlations between the  concomitants $(k=1,\ldots,4)$ and  the \BL{Malignant Tumours},  we compute the average weight matrix as  
\begin{eqnarray*}
\bar{{\bf D}}_1^{[2]}=\left[\begin{array}{ccccc} 
 0.4766 & 0.4766 & 0.04680 & 0.00000 & 0.0000  \\
 0.4766 & 0.4766 & 0.00000 & 0.04680 & 0.0000 \\
 0.0468 & 0.0000 & 0.47685 & 0.47635 & 0.0000 \\
 0.0000 & 0.0468 & 0.00000 & 0.22710 & 0.7261 \\
 0.0000 & 0.0000 & 0.47635 & 0.24975 & 0.2739 \\
  \end{array} \right],
\end{eqnarray*}  
which will be used to identify the unit in the set ${\bf U}^{[2]}_1$ for full quantification. Since our goal here is to measure the second judgment order statistic $Y_{[2]1}$, we focus on the second column of $\bar{{\bf D}}_1^{[2]}$. We observe that units $\{u_{11},u_{21}\}$ both have the maximum chance to be the second order statistic in this set. Finally, having the higher concentration, the first unit $u_{11}$ is selected as the second judgment  order statistic $Y_{[2]1}$ from  ${\bf U}^{[2]}_1$ for full measurements. The observed data in this example is given by 
\[
(Y_{[2]1},{\lo}^{[2]}_1)=\left(Y_{[2]1},\{0.4766 , 0.4766 , 0.04680 , 0 , 0\}\right),
\]
where ${\lo}^{[2]}_{1}$ is the \cc{first} row of the matrix $\bar{{\bf D}}_1^{[2]}$ (the row corresponding to \cc{$u_{11}$}). Note that we also observe the concomitant values associated with the selected unit that can be used in the estimation of the probability of having malignant breast tumours  using some  regression models such as logistic regression. 

\section{Statistical procedures } \label{sec:3}

\subsection{Multi-concomitant \BL{PROS} estimator}

Let ${\bf Y}_{mp}=\{(Y_{[r]i},{\lo}^{[r]}_{i});r=1,\ldots,H;i=1,\ldots,\cc{n}\}$
 denote the multi-concomitant PROS sample of size $\cc{n}H$ with set size $H$ and cycle size $\cc{n}$. Suppose $Y_{[r]i}$ is
 the quantified $r$-th judgment order statistic \BL{and $\lo_i^{[r]}=({\bar\omega}_i^{[r,1]},\ldots,{\bar\omega}_i^{[r,H]})$ is  its corresponding weight vector such that $0\le {\bar\omega}_i^{[r,h]} \le 1$ and $\sum_{h=1}^{H}{\bar\omega}_i^{[r,h]}=1$.}
 To exploit the tie-structure in the estimation of  the population proportion $p$, the quantified statistics are prorated to the judgment ranks, using the strength-of-weight probability vector. Due to \ww{\cite{ozturk2013estimation}}, the population proportion estimation under the multi-concomitant PROS data is calculated by  
\begin{eqnarray}\label{est-mcp}
 {\hat p}_{mp}= \frac{1}{H}\sum_{\BL{h}=1}^{H} \frac{\BL{\sum_{r=1}^{H}}\sum_{i=1}^{n} 
 {\bar\omega}_i^{[r,\BL{h}]} Y_{[r]i}}{\frac{1}{n\BL{H}}\BL{\sum_{r=1}^{H}}\sum_{i=1}^{n}
 {\bar\omega}_i^{[r,\BL{h}]}} 
 = \BL{\sum_{r=1}^{H}}\sum_{i=1}^{n} {\tilde\omega}^{[r]}_{i} Y_{[r]i},
 \end{eqnarray}
where 
\[
 {\tilde\omega}^{[r]}_{i}=\frac{1}{H}\sum_{\BL{h}=1}^{H} \frac{ {\bar\omega}_i^{[r,\BL{h}]} }{\frac{1}{n\BL{H}}
 \BL{\sum_{r=1}^{H}}\sum_{i=1}^{n}{\bar\omega}_i^{[r,\BL{h}]}}.
\] 
Note that ${\bar\omega}_i^{[r,\BL{h}]} Y_{[r]i}$ can be interpreted as the allocation of $Y_{[r]i}$ to the \BL{$h$}-th judgment rank, proportional to the strength of  the agreement probability that $h$ is the true rank of  $Y_{[r]i}\, (h=1,\ldots,H)$. Note that population proportion estimator using \ww{one-concomitant RSS  (\citealp{terpstra2004concomitant}) can be obtained   as an special case of  \eqref{est-mcp} using a single concomitant with no tie-structure.} 

\subsection{RSS-based Logistic regression estimator}

\ww{\cite{chen2005ranked}} used multiple concomitants to obtain   an RSS-based estimator of the  population proportion using a 
   logistic regression model on concomitants $\bf X$  as follows
\begin{eqnarray} \label{log2}
p=\frac{\exp(\beta_0+{\bf\beta}^{\top} {\bf X})}{1+\exp(\beta_0+{\bf{\beta}}^{\top} {\bf X})},
\end{eqnarray}
where $p$ is the corresponding probability of success, $\beta_0$ is the intercept parameter and $\bf\beta$ is the vector of slope parameters. 
\ww{It is  assumed that there is a "training data set"  consisting of  the values of the  concomitants as well as the \BL{response} variable. \cite{chen2005ranked} requires the training data set  to estimate the  parameters of the 
 logistic regression model  in  \eqref{log2} (i.e., $\beta_0$ and ${\bf\beta}$).} 
Let ${\bf X}_r$ denote the vector of   concomitants associated with   the $r$-th individual  in a set of size $H$.  \ww{\cite{chen2005ranked}}   estimates the probability of success ${\hat p}_r$ $(r=1,\ldots, H)$ based on the   fitted logistic regression model and    use them for   ranking the sampling units with binary response variable $Y$ to  obtain  an RSS  sample of size $\cc{n}H$ given by  \cc{$\{Y_{[r]i},r=1,\ldots,H;i=1,\ldots,n\}$}.  
Finally,  they propose an  RSS-based  estimator of the  population proportion $p$ as follow
 \begin{eqnarray}
{\hat p}_{l}=\frac{1}{\cc{n}H}\sum_{r=1}^{H} \sum_{i=1}^{\cc{n}} Y_{[r]i}.
\end{eqnarray}
We refer to this method as the  multi-concomitant RSS-based logistic regression method  for  estimating  $p$. 

 \subsection{Standard deviation  reduction}
 To evaluate the performance of the multi-concomitant PROS estimator ${\hat p}_{mp}$,
 \ww{ we compare the average, standard deviation (SD) as well as  the SD reduction of ${\hat p}_{mp}$ with those of the counterpart estimators under simple random sampling (SRS), one-concomitant RSS (\citealp{terpstra2004concomitant})  and multi-concomitant  RSS-based   logistic regression   methods (\citealp{chen2005ranked}).} 
 Let $\bar{\hat{p}}_{mp}=\frac{1}{J}\sum_{j=1}^{J} \hat{p}_{mp}^{j}$, where  $\hat{p}_{mp}^{j}$ is the multi-concomitant PROS estimate of $p$ obtained from  the  $j$-th replicate, $j=1, \ldots, J$.
 The SD of $\hat{p}_{mp}$ is  then computed by 
\begin{eqnarray} \label{sd-mp}
\text{SD}(\hat {p}_{mp}) = \left\{
\frac{1}{J-1} \sum_{j=1}^{J} (\hat{ p}_{mp}^{j}-\bar{\hat{p}}_{mp})^{2}
\right\}^{1/2}.
\end{eqnarray}
Similar  to \ww{\cite{chen2005ranked}}, 
since $p$ is known, using the finite population correction (population size  is $N'$), we calculate the SD of the population proportion estimator based on a SRS sample of size $m$ as follows 
\begin{eqnarray} \label{sd-srs}
\text{SD}({\hat p}_{srs})= \left\{
\frac{N'-m}{N'-1} \times \frac{p(1-p)}{m}
\right\}^{1/2}.
\end{eqnarray}
From \eqref{sd-mp} and \eqref{sd-srs}, the percentage of  reduction in the sample SD (SD reduction) by  using ${\hat p}_{mp}$ instead of ${\hat p}_{srs}$ is computed by 
\begin{eqnarray*} \label{sd-red}
\text{SD reduction($\hat {p}_{mp}$, $\hat{p}_{srs}$)}=\left(
1-\frac{\text{SD}({\hat p}_{srs})}{\text{SD}({\hat p}_{mp})}
\right)\times 100.
\end{eqnarray*}
The average and SD and SD reduction measures can similarly be obtained for  other estimation procedures. 

\section{Breast cancer data analysis} \label{sec:4}
In this section, we use the WBCD as the underlying population of interest, to evaluate the performance of ${\hat p}_{mp}$ for  estimating the proportion of patients with malignant breast cancer tumour and compare it with its counterparts under SRS, one-concomitant RSS and  multi-concomitant  RSS-based logistic regression methods. The WBCD consists of 699 subjects in where the tumour of $266$ subjects have been identified as malignant. For this study,  the response  \BL{variable is the   "Malignant Tumours"}  that follows a Bernoulli distribution  with probability of success $p=0.3499$.
Through two numerical studies, we first evaluate the performance of ${\hat p}_{mp}$  for  estimating the proportion of patients with  malignant  breast tumours compared with its  counterparts  under SRS, one-concomitant RSS and multi-concomitant RSS-based logistic regression methods. Then,  we investigate the effect of tie-structures on the performance of ${\hat p}_{mp}$.

\newpage
\noindent \textbf{Study 1: One-concomitant  and multi-concomitant RSS procedures}

\ww{In this study, we evaluate the performance of the population proportion estimators under different sampling procedures. We focus on   SRS estimator of $p$, one-concomitant RSS-based    estimator \ww{(\citealp{terpstra2004concomitant})},  multi-concomitant   RSS-based logistic regression  estimator \ww{(\citealp{chen2005ranked})}, and our proposed   multi-concomitant PROS estimator   given in \eqref{est-mcp}.} The sample size (i.e., number of measured units) is fixed at $54$ for each setting. 
To calculate the average, SD and SD reduction measures, each estimation procedure is replicated $50,000$ times. 
Since the population proportion is $p=0.3499$ (with population size $N'=699$), the SD of ${\hat p}_{srs}$  based on a SRS  without replacement  of  size $54$ is given by 
\begin{eqnarray} \label{srs-sd}
\text{SD}(\hat{p}_{srs})=\sqrt{\frac{699-54}{699-1}\times\frac{0.3499(1-0.3499)}{54}}=0.0623.
\end{eqnarray}
To evaluate the performance of one-concomitant RSS-based  estimator, we select five single concomitants for the  ranking process of the RSS technique.  These include   Bare Nuclei, Uniformity of Cell Shape and Uniformity of Cell Size which have the highest correlations with  the \BL{Malignant Tumours}, as well as the 
Normal Nuclei,  the \BL{Subject ID and Independent Covariate}  which\BL{, respectively,} have  intermediate, very low  \BL{and almost zero }correlations with \BL{Malignant Tumours} to study  the effect of different ranking abilities on the performance of one-concomitant  RSS-based  estimator of $p$. The RSS design  with set sizes $H=3,6$ and $9$ are then  applied to obtain the RSS estimates under these five single concomitants.
{\small
\begin{table}[h!]
\caption{\small{Averages and SDs (in bracket) of 50,000 estimates of the malignant proportion.}}
\vspace{0.3cm} 
\centering 
\small{\begin{tabular}{lcccc}\hline\hline
   & \multicolumn{4}{c}{Sampling Designs} \\ 
                  \cline{2-5} 
Concomitants             & SRS  & RSS with H=3 &  RSS with H=6  & RSS with H=9 \\ \hline
Bare Nuclei              & 0.4466(0.0623) & 0.3296(0.0557) & 0.3130(0.0487) & 0.3207(0.0442) \\ 
Uniformity of Cell Shape & 0.4473(0.0623) & 0.3281(0.0547) & 0.3049(0.0456) & 0.3069(0.0389) \\ 
Uniformity of Cell Size  & 0.4472(0.0623) & 0.3262(0.0540) & 0.3033(0.0449) & 0.3076(0.0384) \\ 
Normal Nucleoli          & 0.4468(0.0623) & 0.3369(0.0573) & 0.3274(0.0516) & 0.3279(0.0477) \\ 
\BL{Subject ID}          & 0.4466(0.0623) & 0.3558(0.0596) & 0.3549(0.0576) & 0.3524(0.0555) \\ 
\BL{Independent Covariate}& \BL{0.4472(0.0623)} & \BL{0.3408(0.0588)} & \BL{0.3111(0.0558)} & \BL{0.3208(0.0547) } \\\hline
\end{tabular}
\label{tab:1-ave}}
\end{table}
}

 Tables \ref{tab:1-ave} and \ref{tab:1-sdr} provide the average, SD and SD reduction values of these five one-concomitant RSS estimates obtained through $50,000$ replicates.
Our results show that RSS estimators based on single concomitant perform \cc{ well} in the estimation of $p$ for all the set sizes compared with their SRS counterparts.
Focusing on Uniformity of Cell Shape as the ranking concomitant, for instance, the sample SD of the RSS estimator compared with that of SRS estimator reduces from $13.34\%$ for set size $H=3$ to $38.39\%$  for set size $H=9$. 
Due to the fact that the \BL{Subject ID and Independent Covariate} have the smallest correlations with the \BL{Malignant Tumours}, as expected, the percent reduction in SD of one-concomitant RSS-based  estimator under \BL{these concomitants are} low for all the set sizes. However,  as the set size increases the  SD reductions  of  the  one-concomitant  RSS-based estimators increases.


{\small
\begin{table}[h!]
\caption{\small{Percent SD reductions of RSS estimators relative to SRS estimator of the malignant proportion.}}
\vspace{0.3cm} 
\centering 
\small{\begin{tabular}{lccc}\hline\hline
   & \multicolumn{3}{c}{RSS with set size} \\ 
                  \cline{2-4} 
Concomitants    &  H=3 &   H=6  &  H=9 \\ \hline
Bare Nuclei             & 10.5713 & 21.9384 & 29.1293 \\ 
Uniformity of Cell Shape&  12.2834 & 26.8765 & 37.5440 \\ 
Uniformity of Cell Size &  13.3495 & 27.9003 & 38.3916 \\ 
Normal Nucleoli         &  8.1039 & 17.2002 & 23.4274 \\ 
\BL{Subject ID}         &  4.3091 & 7.5649 & 11.0161 \\ 
\BL{Independent Covariate}& \BL{5.6549} & \BL{10.4947} & \BL{12.2978}  \\\hline

\end{tabular}
\label{tab:1-sdr}}
\end{table}
}

Now, \ww{we consider estimation of the  population proportion using multi-concomitant   PROS   and  RSS-based  logistic regression estimators (\citealp{chen2005ranked}). We then compare them with the results obtained under one-concomitant RSS estimator (\citealp{terpstra2004concomitant}).} We consider nine  ranking  models corresponding to different choices of concomitants,  as described in Table \ref{tab:1-rm}. Models 1, 2 and  9 focus on concomitants which have the highest (Bare Nuclei) and the lowest (\BL{Subject ID and Independent Covariate})  correlation coefficients with  the \BL{Malignant Tumours}, respectively.
These  models will help us to simultaneously explain the effect of different ranking potentials on the performance of the malignant proportion estimators. \ww{We can also compare them with the results presented in Tables \ref{tab:1-ave} and \ref{tab:1-sdr} for  one-concomitant RSS method.} 
Models 3 and  4 are \cc{two-concomitant} based ranking models whose logistic regression models are significant. These models allow us  to compare the performance of multi-concomitant PROS with that of the RSS-based  logistic regression model  in the estimation of malignant proportion. 
{\small
\begin{table}[h!]
\caption{\small{Various ranking models using various concomitants.}}
\vspace{0.3cm} 
\centering 
\small{\begin{tabular}{clc  c}\hline\hline
Ranking       & Explanatory  & Degree of association with   &  Significancy   \\
 ~Models      &  ~~~variables &Malignant Tumour  & of the Model at $\alpha=0.05$ \\ \hline
Model 1     &  Bare Nuclei    & Very Strong & Yes\\
Model 2     &  \BL{Subject ID} & Very Low &No \\
Model 3     &  Bare Nuclei, Normal Nucleoli & Very Strong, Intermediate&Yes \\
Model 4     &  Bare Nuclei, Single Epithelial Cell Size & Very Strong, Intermediate &Yes \\
Model 5     &  Bare Nuclei, Uniformity of Cell Size,&  Very Strong, Very Strong, & No \\[-1ex]
            &  Uniformity of Cell Shape & Very Strong & \\
Model 6     &  Bare Nuclei, Uniformity of Cell Size, &  Very Strong, Very Strong, & No \\[-1ex]
            &  \BL{Subject ID}& Very Low & \\
Model 7     &  Bare Nuclei, Normal Nucleoli, & Very Strong, Intermediate, & No \\[-1ex]
            &  Clump Thickness, Bland Chromatin&  Intermediate, Intermediate &\\
Model 8     &  Normal Nucleoli, Clump Thickness, &   Intermediate, Intermediate, & Yes  \\[-1ex]
            &  Bland Chromatin & Intermediate &  \\
\BL{Model 9}&  \BL{Independent Covariate}&  No association  & No \\\hline          
\end{tabular}
\label{tab:1-rm}}
\end{table}
}

Model 5 and 6 are two ranking models using three concomitants for ranking purposes. 
 Model 5 includes three concomitants with the highest correlations with the  \BL{Malignant Tumours}, while Model 6 involves two best ranking variables along with the worst ranking variable. 
These models explain  the effect of the worst ranking variable in the presence of the best ranking variables.
Lastly, Model 7 consists of the best ranking variable along with the 3 ranking variables with intermediate correlations with the  \BL{Malignant Tumours} 
 while Model 8 only consists of  three concomitants with  intermediate ranking abilities  that are used in  Model 7. This assists us to explain how much precision is lost by using intermediate ranking variables when we do not have access to the best ranking variable.

{\small
\begin{table}[h!]
\caption{\small{\BL{Averages (AV), SDs (SD) and percent SD reductions (PSR) of 50,000  ${\hat p}_{mp}$ using various ranking models with set sizes $H=\{3,6,9\}$.}} }
\vspace{0.3cm} 
\centering 
\small{\BL{\begin{tabular}{cccccccccccc} \hline
  & \multicolumn{3}{c}{H=3} & & \multicolumn{3}{c}{H=6} & &\multicolumn{3}{c}{H=9} \\
    \cline{2-4}              \cline{6-8}   \cline{10-12} 
 Models     & AV  &  SD  & PSR  & & AV  &  SD  & PSR & & AV  &  SD  & PSR \\ \hline
Model  1 & 0.3437 & 0.0507 & 18.5899 & &0.3303 & 0.0444 & 28.7434 & &0.3334 & 0.0422 & 32.2968 \\ 
Model  2 & 0.3562 & 0.0594 & 4.6454  & &0.3546 & 0.0578 & 7.3142  & &0.3506 & 0.0558 & 10.5121 \\ 
Model  3 & 0.3388 & 0.0495 & 20.6386 & &0.3233 & 0.0405 & 35.0739 & &0.3237 & 0.0361 & 42.0048 \\ 
Model  4 & 0.3336 & 0.0510 & 18.1295 & &0.3143 & 0.0399 & 35.9273 & &0.3130 & 0.0356 & 42.8559 \\ 
Model  5 & 0.3322 & 0.0480 & 23.0078 & &0.3219 & 0.0376 & 39.6046 & &0.3184 & 0.0322 & 48.4163 \\ 
Model  6 & 0.3358 & 0.0487 & 21.8970 & &0.3234 & 0.0367 & 41.1436 & &0.3224 & 0.0334 & 46.3639 \\ 
Model  7 & 0.3282 & 0.0491 & 21.1707 & &0.3181 & 0.0388 & 37.8145 & &0.3182 & 0.0338 & 45.8352 \\ 
Model  8 & 0.3281 & 0.0510 & 18.2574 & &0.3165 & 0.0411 & 34.0816 & &0.3048 & 0.0359 & 42.4212 \\ 
Model  9 & 0.3411 & 0.0591 & 5.2142  & &0.3199 & 0.0569 & 8.6820  & &0.3197 & 0.0547 & 12.2053 \\ 
   \hline
\end{tabular}
\label{tab:1-mcp}}}
\end{table}
}

The PROS samples using multi-concomitants are constructed as described in Section \ref{sec:2}. We use the tie-structure model introduced by \ww{\cite{frey2012nonparametric}}. For instance, focusing on a concomitant, say $X_1$,  we divide $X_1$ with a nonnegative $c$, $X_1/c$, and then round it to the nearest integer. Let $X_1^{*}$ denote the discretized  version of $X_1/c$. Then,  units  with the same discretized value are declared as the tied-ranks   in the set. This discretization   process is considered for all  available concomitants. The multi-concomitant PROS data are finally generated through the combined ranking information from  all available concomitants with  the possibility of  tie-structures.
For more information, see \ww{ \cite{ozturk2013combining} and \cite{frey2012nonparametric}}.

\ww{As mentioned earlier, there are two disadvantages associated with  RSS-based logistic regression  estimation procedure (\citealp{chen2005ranked}).}
 The first disadvantage is that a  `training sample', which requires  the quantification of the \BL{Malignant Tumours},  is needed  to  estimate  the  logistic regression model for the  ranking process involved in RSS sampling; however, the multi-concomitant PROS estimation procedure does not need the training data set.
 To fit the RSS-based  logistic regression model, a `training sample' of size 100 was taken at random from the  WBCD. 
 This training data set  is  used for estimating the probabilities  of success and performing  the  ranking process required for  RSS estimation procedure of \ww{\cite{chen2005ranked}}. 

Under only  the Bare Nuclei ranking variable, ${\hat p}_{mp}$ performs very well in the estimation of the malignant proportion. The sample SD reduction of ${\hat p}_{mp}$ using Model 1 (Table \ref{tab:1-mcp}) accounts for \BL{18.58\%} for $H=3$ and \BL{32.29\%} for $H=9$ while precision gained by the counterpart estimators \BL{for $H=3$ are 
almost 10\% and 29\% for $H=9$}  in Table \ref{tab:1-sdr} and Table \ref{tab:1-log},  respectively.  This indicates the superiority of  ${\hat p}_{mp}$ over its counterparts even  with  one concomitant for ranking. \ww{The relative efficiency (RE) is another valuable measure to compare the performance of the estimation procedures (\citealp{terpstra2004concomitant})}. The relative efficiency  can be considered   as the ratio of variances of two estimators as $RE({\hat p}_{mp},{\hat p}_l)=Var({\hat p}_{mp})/Var({\hat p}_l)$, and  can be interpreted as the ratio of the required  sample sizes  to obtain the same precision for the two estimation procedures, say $N_l=RE({\hat p}_{mp},{\hat p}_l) N_{mp}$,  where $N_l$ and $N_{mp}$ are the total sample sizes of the estimation procedures. From  Tables   \ref{tab:1-mcp} and \ref{tab:1-log}, focusing on the SDs of  ${\hat p}_{mp}$ and ${\hat p}_l$ using Models  3 and 4 \BL{for $H=3$},  ${\hat p}_{l}$ requires samples of sizes \BL{61 and 58}, respectively,  to be as precise as ${\hat p}_{mp}$ with sample size 54 in the estimation of the proportion of  malignant tumours.

The second disadvantage associated with the RSS-based  logistic regression model  is the requirement of the logistic regression model assumption.
To show the impact of such  assumption,  we used forward and backward  selection  methods   for all  concomitant-based ranking models introduced in Table \ref{tab:1-rm} as well as  other possible  models based on available concomitants.  
In the case of one-concomitant ranking model, the regression models based  on  Uniformity of Cell Shape, Bland Chromatin, Clump Thickness and Marginal Adhesion were not significant. 
Models 2, 5, 6, 7 and 9 are considered as examples of   ranking  models were   their  corresponding logistic regression models  are  not  significant.

{\small
\begin{table}[h!]
\caption{\small{\BL{Averages (AV), SDs (SD) and percent SD reductions (PSR) of 50,000  ${\hat p}_{l}$ using various logistic regression models for ranking with set sizes $H=\{3,6,9\}$.}}}
\vspace{0.3cm} 
\centering 
\small{\BL{\begin{tabular}{cccccccccccc} \hline
  & \multicolumn{3}{c}{H=3} & & \multicolumn{3}{c}{H=6} & &\multicolumn{3}{c}{H=9} \\
    \cline{2-4}              \cline{6-8}   \cline{10-12} 
 Models      & AV  &  SD  & PSR  & & AV  &  SD  & PSR & & AV  &  SD  & PSR \\ \hline
Model 1  & 0.3297 & 0.0555 & 10.9610 && 0.3131 & 0.0487 & 21.9186 && 0.3205 & 0.0443 & 28.9573 \\ 
Model  2 & 0.3471 & 0.0590 & 5.3167  && 0.3458 & 0.0571 & 8.4344  && 0.3455 & 0.0568 & 8.8135 \\ 
Model  3 & 0.3295 & 0.0545 & 12.4943 && 0.3109 & 0.0461 & 26.0168 && 0.3182 & 0.0405 & 35.0400 \\ 
Model  4 & 0.3256 & 0.0541 & 13.2606 && 0.3040 & 0.0452 & 27.5200 && 0.3103 & 0.0392 & 37.1236 \\ 
Model  5 & 0.3302 & 0.0559 & 10.3698 && 0.3131 & 0.0490 & 21.3125 && 0.3206 & 0.0442 & 29.1343 \\ 
Model  6 & 0.3245 & 0.0538 & 13.6537 && 0.3039 & 0.0454 & 27.1731 && 0.3072 & 0.0395 & 36.5843 \\ 
Model  7 & 0.3260 & 0.0539 & 13.5306 && 0.3047 & 0.0450 & 27.7898 && 0.3077 & 0.0382 & 38.7863 \\ 
Model  8 & 0.3327 & 0.0544 & 12.7289 && 0.3114 & 0.0455 & 26.9282 && 0.3173 & 0.0389 & 37.5177 \\ 
Model  9 & 0.3406 & 0.0589 & 5.5015  && 0.3112 & 0.0556 & 10.7490 && 0.3209 & 0.0542 & 13.0245 \\ 
   \hline
\end{tabular}
\label{tab:1-log}}}
\end{table}
}

From Table \ref{tab:1-mcp}, focusing on Models 1 and 5, we take the advantage of  multi-concomitant PROS sampling schemes and improve 
the precision  of ${\hat p}_{mp}$ from \BL{$18.58\%$ for $H=3$ and  $32.29\%$ for $H=9$ } under Model 1 (using only the  Bare-Nuclei concomitant) to \BL{$23.00\%$ for $H=3$ and  $48.41\%$ for $H=9$ } under Model 5 in which we benefit from  the full ranking information of  the best three ranking variables in the estimation of the   proportion of patients with malignant tumours.
Considering Models  5 and 6, it is seen  that the precision gain of ${\hat p}_{mp}$ under Model 6 is slightly less than that under Model 5. This illustrates the effect of the bad ranker  on  the estimation; however,  the discrepancy is not too large. This may reflect the benefit of  multi-concomitant PROS procedure in which the good rankers  downside  the impact of a bad ranker. Focusing on Models 1, 7 and 8 \BL{under $H=3$}, it is apparent that the percentage of  SD reduction increases from \BL{$18.58\%$} under Model 1 to 
\BL{$21.17\%$} under Model 7. This indicates that the excellence of ${\hat p}_{mp}$ based on only the  Bare Nuclei
concomitant can be increased by using more intermediate ranking concomitants.
Comparing Models 7 and  8, it is seen that using highly correlated concomitants for the ranking process plays an  important role in the accuracy of  ${\hat p}_{mp}$ in the estimation of  the proportion of patients with malignant tumours. Moreover, we see from Table \ref{tab:1-mcp} that precision gain of ${\hat p}_{mp}$ is almost the same  under Models 1 and  8. This  is  interesting as  if there is no highly correlated concomitant  associated with the  \BL{Malignant Tumours} for the  ranking process, one can still estimate the  proportion of malignant breast tumours  reasonably well   using multiple cytological characteristics having intermediate correlations with the  \BL{Malignant Tumours}.
\noindent \textbf{Study 2: Analysis of the tie-structures}

Here we study the properties of  $\hat{p}_{mp}$  for estimating  the population proportion under different  tie-structures. To this end, we compare the performance of $\hat{p}_{mp}$ using different multi-concomitant PROS samples of fixed size $\BL{n}H=54$  with set sizes $H \in \{2,3,6\}$ and various  tie-structures associated \cc{ with}  $\BL{c}\in\{1,1.5,3,4\}$. \ww{The tie-structures in  PROS samples are constructed as described in Study 1 using discritization formula (\citealp{frey2012nonparametric})}. 

{\small
\begin{table}[h!]
\caption{\small{Results of  50,000 replicates of ${\hat p}_{mp}$ when no tie-structure (i.e., $\BL{c}=1$ for all set sizes $H=\{2,3,6\}$) is allowed in sampling.}}
\vspace{0.3cm} 
\centering 
\small{\begin{tabular}{lccccccc}\hline\hline
Ranking Models  & $H$    & Average &  SD  & SD \rr{Reduction} & Lower \rr{Bound} & Upper \rr{Bound} & Length of CI \\ \hline
Model 1  &  2 & 0.3378 & 0.0576 & 7.6567 & 0.2440 & 0.4333 & 0.1893 \\ [-1ex]
         & 3  & 0.3455 & 0.0513 & 17.6955 & 0.2616 & 0.4302 & 0.1686 \\[-1ex]
         & 6 & 0.3298 & 0.0451 & 27.5769 & 0.2560 & 0.4044 & 0.1484  \\ 
     
Model 2  & 2 & 0.3611 & 0.0601 & 3.6514 & 0.2590 & 0.4630 & 0.2040 \\[-1ex]
         & 3 & 0.3565 & 0.0601 & 3.6224 & 0.2592 & 0.4619 & 0.2028 \\[-1ex]
         & 6 & 0.3545 & 0.0576 & 7.5525 & 0.2593 & 0.4463 & 0.1870 \\
      
Model $5^*$ &  2 & 0.3295 & 0.0564 & 9.5027 & 0.2369 & 0.4236 & 0.1868 \\ [-1ex]
            & 3 & 0.3327 & 0.0487 & 21.9011 & 0.2526 & 0.4120 & 0.1594 \\[-1ex] 
            & 6 & 0.3124 & 0.0399 & 36.0475 & 0.2468 & 0.3781 & 0.1313 \\ 
        
Model 3  & 2 & 0.3333 & 0.0568 & 8.9168 & 0.2390 & 0.4276 & 0.1886 \\ [-1ex]
         & 3 & 0.3377 & 0.0497 & 20.2667 & 0.2563 & 0.4196 & 0.1634 \\ [-1ex]
         & 6 & 0.3227 & 0.0410 & 34.2832 & 0.2558 & 0.3902 & 0.1345 \\
         
Model 5    &  2 & 0.3294 & 0.0559 & 10.3764 & 0.2382 & 0.4226 & 0.1845 \\ [-1ex]
           & 3 & 0.3344 & 0.0486 & 22.0833 & 0.2546 & 0.4141 & 0.1595 \\ [-1ex]
           & 6 & 0.3154 & 0.0386 & 38.0544 & 0.2525 & 0.3793 & 0.1268 \\ 
        
Model 8    &  2 & 0.3314 & 0.0571 & 8.3292 & 0.2410 & 0.4262 & 0.1853 \\ [-1ex] 
           &  3 & 0.3338 & 0.0512 & 17.7815 & 0.2489 & 0.4188 & 0.1699 \\[-1ex]
           &  6 & 0.3050 & 0.0414 & 33.5407 & 0.2379 & 0.3739 & 0.1360 \\
           
\BL{Model 9}&\BL{ 2 }&\BL{ 0.3499 }&\BL{ 0.0591 }&\BL{ 5.2576  }&\BL{ 0.2407 }&\BL{ 0.4444 }&\BL{ 0.2037} \\ [-1ex]
            &\BL{ 3 }&\BL{ 0.3406 }&\BL{ 0.0587 }&\BL{ 5.7845  }&\BL{ 0.2222 }&\BL{ 0.4444 }&\BL{ 0.2222} \\ [-1ex]
            &\BL{ 6 }&\BL{ 0.3109 }&\BL{ 0.0555 }&\BL{ 10.8859 }&\BL{ 0.2037 }&\BL{ 0.4074 }&\BL{ 0.2037}  \\\hline
\end{tabular}
\label{tab:2-nts}}
\end{table}
}

%
 \BL{ The tie-structure is selected roughly proportional to  $c=\delta/H$ where  $H$ is set size and $\delta$ is the range for the associated concomitant variable.
This selection approximately assigns  the same number of categories to the subsets  and proposes roughly balanced PROS samples; however,  it should be noted that the final samples may not be necessarily  balanced, since the units are assigned to the subsets based on  different ranking potentials of concomitants.  Although the ordinal concomitants consist of  10 categories, since category 9 is rarely observed compared with other categories, we used $\delta=9$ (instead of 10) in the selection of  tie-structures for these concomitants. 
 More specifically, when the set size $H=2$, the tie-structure is $c=\delta/H=4.5\sim 4$. For set size $H=3$, we use  the tie-structure $c=\delta/H=3$. Similarly $c=\delta/H=1.5$ for the case when $H=6$.}
Also, note that  $\BL{c}=1$ indicates that no-tie structure is created through PROS sampling; however,  we may still \rr{observe} tied-ranks in the estimation procedure because of the nature of the concomitants. 
For  the \BL{Subject ID} we have $\delta=13390977$ and  the tie-structures are selected roughly proportional to $\delta/H$ for  different set sizes $H\in \{2,3,6\}$. In this study, we consider $6$ sets of concomitants (as ranking models) for  the ranking purpose.
 These  include Models 1, 2, 3, 5, 8 and 9 of Table \ref{tab:1-rm} as well as Model $5^*$ using  the  Bare Nuclei and the  Uniformity of Cell Shape as two good ranking variables.
  We obtain ${\hat p}_{mp}$  based on  different multi-concomitant PROS samples  and then the estimation procedures are replicated 50,000 times. 
 We calculate the averages and SD of 50,000 malignant proportion estimates for each setting.
To evaluate the effect of different tie-structures on the performance of ${\hat p}_{mp}$ in the estimation of malignant proportion, the reduction gained in sample SD of ${\hat p}_{mp}$ under each setting is compared with \eqref{srs-sd} under its counterpart based on without replacement SRS sample of size $54$.
 For each setting, we also study $90\%$ non-parametric confidence intervals (CIs) for the population proportion, where  the lower bound (Lower \rr{Bound}) and upper bound (Upper \rr{Bound}) of the intervals are computed as the empirical \cc{5\%} and \cc{95\%} quantiles  of the proportion of  malignant breast tumours    estimates. We also calculate the length of the confidence intervals for each setting as another  measure of performance. Tables \ref{tab:2-nts} and \ref{tab:2-wts} show the results of the estimation procedures under PROS scheme using different tie-structures.
{\small
\begin{table}[h!]
\caption{\small{Results of  50,000 replicates of ${\hat p}_{mp}$ using different tie-structures ($\BL{c}=\{4,3,2\}$ when set size $H=\{2,3,4\}$, respectively) in sampling.}}
\vspace{0.3cm} 
\centering 
\small{\begin{tabular}{lccccccc}\hline\hline
Ranking Models  & $H$    & Average &  SD  & SD \rr{Reduction} & Lower \rr{Bound} & Upper \rr{Bound} & Length of CI \\ \hline
Model 1  &  2 & 0.3341 & 0.0567 & 9.1005 & 0.2413 & 0.4288 & 0.1875 \\ [-1ex] 
         & 3  & 0.3441 & 0.0505 & 18.9653 & 0.2609 & 0.4275 & 0.1665 \\ [-1ex]
         & 6 & 0.3306 & 0.0446 & 28.4310 & 0.2579 & 0.4047 & 0.1468 \\ 
     
Model 2  & 2 & 0.3498 & 0.0608 & 2.4816 & 0.2411 & 0.4446 & 0.2035 \\[-1ex]
         & 3 & 0.3487 & 0.0601 & 3.6394 & 0.2414 & 0.4447 & 0.2034 \\[-1ex] 
         & 6 & 0.3550 & 0.0593 & 4.7922 & 0.2599 & 0.4498 & 0.1898 \\ 
     
Model $5^*$ &  2 & 0.3334 & 0.0555 & 10.9221 & 0.2436 & 0.4259 & 0.1823 \\ [-1ex]
            & 3 & 0.3310 & 0.0489 & 21.5358 & 0.2507 & 0.4113 & 0.1606 \\[-1ex]
            & 6 & 0.3220 & 0.0400 & 35.8855 & 0.2568 & 0.3885 & 0.1317 \\ 
        
Model 3  & 2 & 0.3345 & 0.0566 & 9.2023 & 0.2424 & 0.4284 & 0.1861 \\ [-1ex]
         & 3 & 0.3384 & 0.0492 & 21.0232 & 0.2576 & 0.4194 & 0.1619 \\ [-1ex] 
         & 6 & 0.3234 & 0.0407 & 34.7509 & 0.2567 & 0.3901 & 0.1334 \\  
        
Model 5    &  2 & 0.3319 & 0.0556 & 10.7361 & 0.2407 & 0.4246 & 0.1839 \\  [-1ex]
           & 3 & 0.3318 & 0.0487 & 21.9182 & 0.2524 & 0.4124 & 0.1600 \\  [-1ex]
           & 6 & 0.3216 & 0.0376 & 39.6662 & 0.2606 & 0.3842 & 0.1236 \\ 
         
Model 8    &  2 & 0.3337 & 0.0571 & 8.3607 & 0.2425 & 0.4282 & 0.1857 \\  [-1ex]
           &  3 & 0.3278 & 0.0509 & 18.4197 & 0.2456 & 0.4114 & 0.1659 \\  [-1ex]
           &  6 & 0.3166 & 0.0410 & 34.1925 & 0.2493 & 0.3841 & 0.1348 \\ 
           
\BL{Model 9} &\BL{ 2 }&\BL{ 0.3468 }&\BL{ 0.0596 }&\BL{ 4.3587 }&\BL{ 0.2521 }&\BL{ 0.4451 }&\BL{ 0.1930} \\[-1ex] 
             &\BL{ 3 }&\BL{ 0.3411 }&\BL{ 0.0592 }&\BL{ 5.0599 }&\BL{ 0.2423 }&\BL{ 0.4416 }&\BL{ 0.1993} \\[-1ex] 
             &\BL{ 6 }&\BL{ 0.3196 }&\BL{ 0.0569 }&\BL{ 8.7842 }&\BL{ 0.2258 }&\BL{ 0.4133 }&\BL{ 0.1874} \\\hline 
                   
\end{tabular}
\label{tab:2-wts}}
\end{table}
}

From Tables \ref{tab:2-nts} and \ref{tab:2-wts}, we observe that  when the set size increases from 2 to 6,  
${\hat p}_{mp}$ performs  better  in the estimation of the proportion of patients with malignant breast tumours,   resulting in  increase in the SD reductions  and decrease in  the length of the CIs. 
Comparing Tables \ref{tab:2-nts} and \ref{tab:2-wts}, it is apparent that appropriate selection of tie-structures improves   the precision of  ${\hat p}_{mp}$ in the estimation of malignant proportion. 
 As noted earlier,  using   suitable tie-structures for  one-concomitant RSS-based  estimator, one can make better inference about the determination of malignancy  of breast tumours than  the one based on the estimation procedure  proposed by \ww{\cite{terpstra2004concomitant} }.
\section{Summary and concluding remarks} \label{sec:5}
In many medical studies,   measuring the variable of interest is costly, time consuming or difficult,  but a small number of sampling units can be ranked  easily using some easy to obtain  concomitants and this can be done at little cost.
In these situations, rank-based sampling designs such as RSS and PROS  sampling techniques   can be efficiently employed to  obtain more representative samples from the underlying population and make  better inference about the parameter  of interest. 
 In this paper, we investigate  the properties of PROS sampling design  with  multiple concomitants for estimating the proportion of patient with malignant tumours in a breast cancer study using  the Wisconsin Breast Cancer Data (WBCD) as the population of interest.   In this application,  the discrimination between the malignant and benign breast tumours requires a comprehensive biopsy procedure which is often time consuming and costly.  However, there are nine visually assessed cytological characteristics that are usually used to  more accurately diagnose  the breast cancer status. These concomitants can be used to obtain better samples from the underlying population using multi-concomitant PROS sampling technique and possibly better estimate  the population proportion. 
To show this,  we proposed an estimator of the population proportion using multi-concomitant PROS sample and,  through extensive numerical studies,  investigated  the effect of different ranking potentials of the concomitants (i.e., good, intermediate and bad)  on the performance of this estimator compared with  its counterparts under SRS, one-concomitant RSS and multi-concomitant RSS-based logistic regression methods.  
Numerical analysis shows that multi-concomitant PROS estimator performs very well compared with its   SRS  counterpart and those  proposed  by  \ww{\cite{terpstra2004concomitant}} as well as \ww{\cite{chen2005ranked}}. 
Unlike the estimator of \ww{ \cite{terpstra2004concomitant}} which is  restricted to only  one concomitant for  the  ranking process, multi-concomitant PROS estimator takes the  full benefit of multiple concomitants and provides significant improvement  in the estimation of proportion.
Although the RSS-based  logistic regression  method of \ww{\cite{chen2005ranked}} uses  multiple concomitants in the estimation of $p$, their estimation method  requires the  logistic regression modeling assumptions. In the absence of such strong modeling assumptions, it is not clear how the extra information should be incorporated in an efficient way.
   In  the WBCD, through different examples, we illustrated that such assumptions are not satisfied even if we use concomitants  that are   highly correlated   with the \BL{Malignant Tumours}. 
Our proposed method can efficiently incorporate as many concomitants as available into the estimation, regardless of such modeling assumptions.
Another advantage of our method is the simple form of our estimator which is simply the weighted average of the PROS sample estimates.
In addition,  the methods proposed in  \ww{\cite{terpstra2004concomitant}} and \ww{\cite{chen2005ranked}} do not allow to declare ties in  ranks in the ranking process of RSS technique. This is not   realistic, in particular for the analysis of the WBCD,  where all the cytological concomitants are \BL{ordinal} variables taking on values between 1 to 10. Multi-concomitant PROS sampling design, through partial ranking, not only eliminates this  restriction but also   results in more accurate estimators. It is worth mentioning that  the proposed methodology  in this paper can be applied to other medical studies as well.
 
%
%
\section*{Acknowledgements}
\jj{Armin Hatefi would like to acknowledge} the University of Manitoba Graduate Fellowship (UMGF) and Manitoba Graduate Scholarship (MGS). Mohammad Jafari Jozani acknowledges the partial  support of  Natural Sciences and Engineering Research Council (NSERC) of Canada. \jj{The authors thank UCI machine learning repository for the assistance received by using the repository.}
\nocite{*}
\bibliographystyle{cbe}


%
%
\end{document}